\documentclass[12pt]{iopart}

\usepackage{graphicx}
\usepackage{bm}
\usepackage{iopams}

\newcommand{\mysize}{8cm}

\newcommand{\mun}[1]{\,\mathrm{#1}}   
\newcommand{\Omcminv}{$\,(\Omega\mathrm{cm})^{-1}$}

\newcommand{\TTF}{TTF--TCNQ}
\newcommand{\THH}{T_H}
\newcommand{\TL}{T_L}
\newcommand{\TI}{T_I}

\newcommand{\mv}[1]{\mathbf{#1}}               
\newcommand{\parl}[2]{\mv{#1}\|\mv{#2}}

\newcommand{\mva}{\mv{a}}   \newcommand{\va}{$\mva$}
\newcommand{\mvb}{\mv{b}}   \newcommand{\vb}{$\mvb$}
\newcommand{\mvc}{\mv{c}}   \newcommand{\vc}{$\mvc$}

\newcommand{\mja}{\parl{j}{a}}          \newcommand{\ja}{$\mja$}
\newcommand{\mjb}{\parl{j}{b}}         \newcommand{\jb}{$\mjb$}

\newcommand{\mBa}{\parl{B}{a}}    \newcommand{\Ba}{$\mBa$}
\newcommand{\mBc}{\parl{B}{c}}    \newcommand{\Bc}{$\mBc$}

\newcommand{\rhoa}{\rho_\mva}
\newcommand{\rhob}{\rho_\mvb}

\newcommand{\siga}{\sigma_\mva}
\newcommand{\sigb}{\sigma_\mvb}
\newcommand{\sigc}{\sigma_\mvc}

\begin{document}

\title[The Hall effect in the organic conductor TTF-TCNQ]{The Hall effect in the organic conductor TTF-TCNQ: Choice of geometry for accurate measurements of highly anisotropic system}

\author{E Tafra$^1$, M \v Culo$^2$, M Basleti\'c$^1$, B Korin-Hamzi\'c$^2$, A Hamzi\'c$^1$ and C S Jacobsen$^3$}
    \address{$^1$Department of Physics, Faculty of Science, University of Zagreb, P.O.Box 331, HR-10002 Zagreb, Croatia}
    \address{$^2$Institute of Physics, P.O.Box 304, HR-10001 Zagreb, Croatia}
    \address{$^3$Department of Physics, Technical University of Denmark, DK-2800 Lyngby, Denmark}
    \ead{etafra@phy.hr}

\begin{abstract}
We have measured the Hall effect on recently synthesized single crystals of quasi-one-dimensional organic conductor TTF-TCNQ, a well known charge transfer complex that has two kinds of conductive stacks: the donor (TTF) and the acceptor (TCNQ) chains.  The measurements were performed in the temperature interval $30\mun{K}<T<300\mun{K}$ and for several different magnetic field and current directions through the crystal. By applying the equivalent isotropic sample (EIS) approach, we have demonstrated the importance of the choice of optimal geometry for accurate Hall effect measurements. Our results show, contrary to past belief, that the Hall coefficient  does not depend on the geometry of measurements and that the Hall coefficient value is around zero in high temperature region ($T > 150$ K), implying that there is no dominance of either TTF or TCNQ chain. At lower temperatures, our measurements clearly prove that all three phase transitions of TTF-TCNQ could be identified from Hall effect measurements.
\end{abstract}


\submitto{\JPCM}

\maketitle

\section{Introduction}

Being the first realization of an organic metal, the quasi one-dimensional molecular crystal \TTF\ (tetrathiafulvalene--tetracyanoquinodimethane) has been studied thoroughly for more than thirty years \cite{R01,R02,R03,R04}. The planar molecules of \TTF\ form segregated stacks in a plane-to-plane manner and the molecular $\pi$--orbitals interact preferably along the stacking direction (crystallographic \vb\ direction of the monoclinic structure), leaving only weak interactions in the perpendicular \va\ and \vc\ crystallographic directions. Due to the formation of linear molecule stacks in the crystal structure and an electronic charge transfer from cationic (TTF) to anionic (TCNQ) complexes, both types of stacks are metallic: TTF column is hole--conducting whereas TCNQ column is electronic--conducting. As a consequence, \TTF\ displays strongly anisotropic metallic conductivity with  $\sigb > \sigc > \siga$ in a wide temperature range down to about $60\mun{K}$, below which a cascade of phase transitions starts and destroys the metallic character progressively.

The noteworthy renewed interest for \TTF\ started recently as it was the first material for which ARPES positively identified 1D single-band Hubbard model spectral features \cite{R05,R06}. Based on these findings the spectral behaviour of \TTF\ was interpreted as an evidence for spin-charge separation, signaling a breakdown of the Fermi liquid quasi--particle picture and leading to the appearance of a new state commonly referred to as one--dimensional quantum many--body system known as Luttinger liquid. However, some quantitative values for different parameters are not yet fully available \cite{R07a,R07b,R07c}.

It is worth pointing out that although the electrical properties of \TTF\ have been studied intensively, the investigations of one basic magnetotransport property, the Hall effect, have not been unambiguously completed. The only published Hall effect measurements of \TTF\ date back to 1977. They were performed in the high temperature metallic region: dc Hall effect measurements in the metallic phase above the phase transitions (for \jb, \Ba) \cite{R08}, and microwave measurements of the Hall mobility at room temperature (for \jb, \Bc) \cite{R09}. However, the obtained results vary in their sign and values. The average value of the Hall coefficient, obtained at room temperature from the dc Hall effect measurements, is negative and approximately consistent with estimates of the electron density implying that the Hall effect and conductivity are dominated by the TCNQ chains. On the other hand, the positive Hall mobility at room temperature, obtained from microwave Hall effect measurements, was interpreted as indication that the carriers in the TCNQ chain relax more rapidly. A direct comparison between those results could not be done not only because the quality of the used samples was most probably not the same, but also because the magnetic field orientations were different in the two experiments. The foreseen measurements in the same geometry (to compare the data and to resolve the question of the importance of the geometry conditions) were not performed. Our aim has been to perform new and systematic Hall effect study on \TTF\ extending them to higher magnetic fields and lower temperatures (i.e.\ below the phase transitions) and using different geometries to investigate the possibility of influence of geometry on the value and sign of Hall coefficient and to try to detect phase transitions,  thereby gain new information.

\section{Experimental details}

\begin{figure}
\centering
 \includegraphics*[width=\mysize]{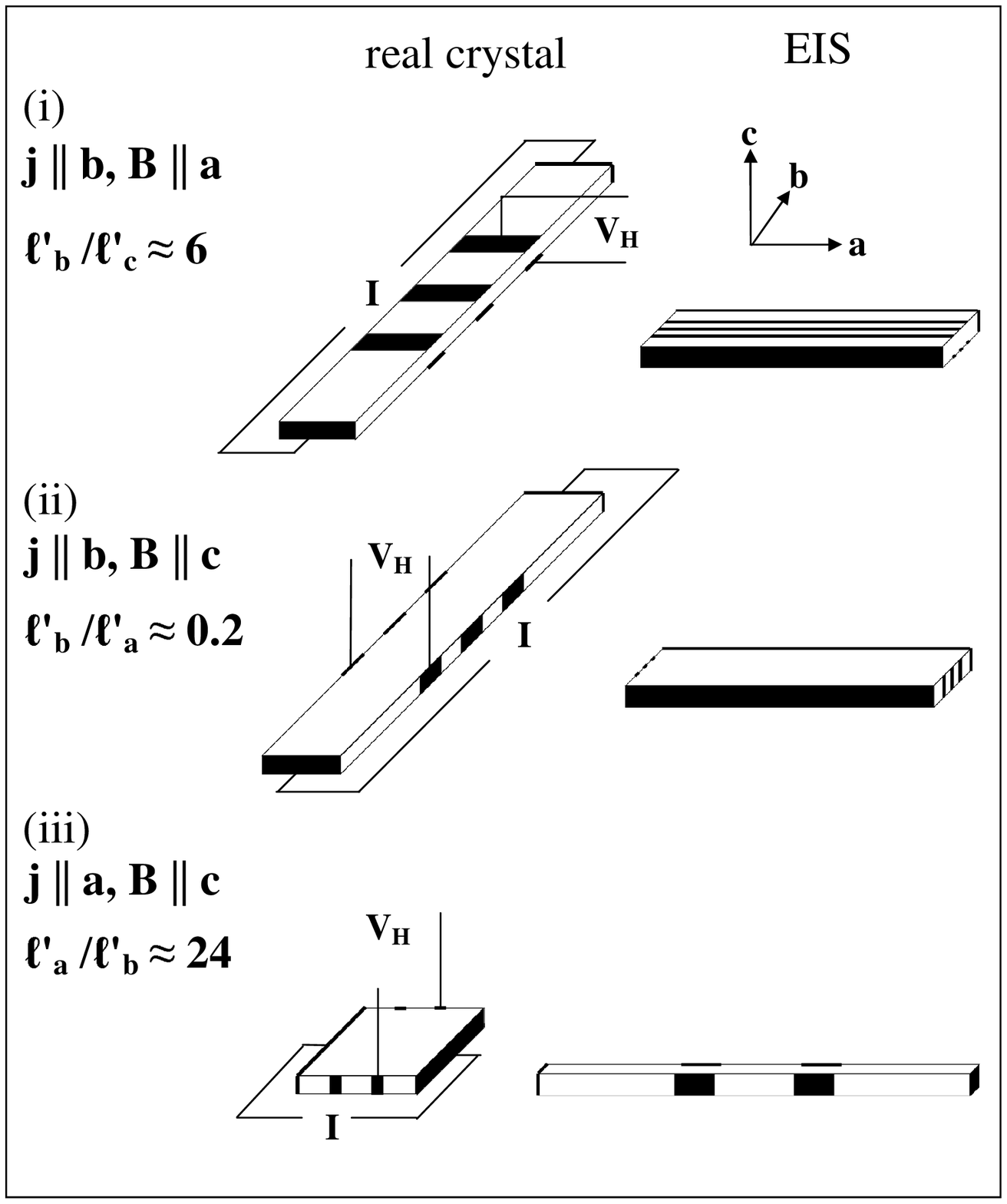}
   \caption{Comparison of the size of the real crystals (left) with equivalent isotropic sample EIS (right), with associated contacts for different geometry used in measurements: (i) \jb, \Ba; (ii) \jb, \Bc; and (iii) \ja, \Bc. The real sample dimensions were $4.7\times 0.7\times 0.08\mun{mm}^3$ for (i) and (ii), and $1.01\times 0.88\times 0.07\mun{mm}^3$ for (iii). We also show the $L'/W'$ ratios, at room temperature, obtained using EIS scaling (see text). In order to be clearly observed, the dimensions in \vc\ direction are doubled in all drawings.} \label{Fig1}
\end{figure}

The measurements of the conductivity and Hall effect were done in the temperature region $30\mun{K}-300\mun{K}$. The Hall effect was measured in 5 and 9~T magnetic fields. All the single crystals used come from the same batch. Their typical configuration is shown on figure \ref{Fig1}: \vb\ direction is the highest conductivity direction, the \va\ direction with the lowest conductivity is perpendicular to \vb\ in the \va--\vb\ plane, and the \vc\ direction with the intermediate conductivity is perpendicular to the \va--\vb\ plane. Our average room temperature conductivity values for $\sigb$ (400\Omcminv), $\sigc$ (5\Omcminv) and $\siga$ (0.5\Omcminv) are in good agreement with previously published data \cite{R03,R04}. In respect to the samples used in previous measurements, we can state that our recently synthesized samples have much larger dimensions and very good defined geometry, show homogeneous current flow, and are of high quality (concerning high $R(300\mun{K})/R(\mathrm{min})$ ratios).

The samples were cooled slowly ($\sim 3\mun{K/h}$) in order to avoid irreversible resistance jumps that are caused by microcracks well known to appear in all organic conductors. Two (or sometimes three, depending on the size of crystal) pairs of Hall contacts and one pair of current contacts were made on the sides of the crystal by evaporating  gold pads to which the $30\,\mu$m gold wires were attached with silver paint. An ac current ($10\,\mu$A to $1\mun{mA}$, $22\mun{Hz}$) was used. For temperatures around and below the phase transitions, a dc technique was also used because of the large resistance increment. Particular care was taken to ensure the temperature stabilization. The Hall voltage was measured at fixed temperatures and in field sweeps from $-B_{\mathrm{max}}$ to $+B_{\mathrm{max}}$ in order to eliminate the possible mixing of the magnetoresistance component. At each temperature the Hall voltage was measured for each pair of Hall contacts to test and/or control the homogeneous current distribution through the sample. However, due to a very steep slope of resistivity in the low temperature region, much better and accurate results were obtained in temperature sweeps in fixed magnetic fields ($-B_{\mathrm{max}}$ and $+B_{\mathrm{max}}$). The Hall voltage $V_H$ was determined as $[V_{xy}(B)-V_{xy}(-B)]/2$ and the Hall coefficient $R_H$ was obtained as $R_H =(V_{H} /IB)t $ ($I$ is the current through the crystal and $t$ is the sample thickness). The linearity of the Hall signal with magnetic field was checked in the whole temperature region investigated.

The Hall effect was measured in several different geometries (i) \jb, \Ba; (ii) \jb, \Bc; and (iii) \ja, \Bc. Moreover, we took care of the concept of the equivalent isotropic sample (EIS) that is crucial for anisotropic conductors \cite{R10}. Generally, for Hall effect measurements it is important to avoid either shorting of the Hall voltage by the current contacts or a non--uniform current distribution.  For isotropic materials this requires that the length/width ($L/W$) ratio obeys the condition $L/W \geq 3$ (where the current is passed along $L$, the Hall voltage develops along $W$ and the magnetic field is along the thickness $t$ of the parallelepiped). For highly anisotropic materials this condition has to be replaced with the more feasible one that takes into account the anisotropy of the conductivity. EIS provides a way of thinking about the current distribution in the anisotropic sample and is realized by a coordinate transformation according to the formulae  $\ell'_i = \ell_i (\sigma/\sigma_i)^{1/2}$, where $ \sigma = (\sigma_1 \sigma_2 \sigma_3)^{1/3}$, $\ell_i$ is the actual sample dimension along the $i^\mathrm{th}$ principal axis of conductivity $ \sigma_i$ and  $\ell'_i$ is EIS dimension along the $i^\mathrm{th}$ axis. Following this notation, we consider that the acceptable geometry for highly anisotropic materials is the one where $L'/W' = L/W(\sigma_{W}/\sigma_{L})^{1/2} \geq 3$.

Figure \ref{Fig1} shows typical crystal configurations with  corresponding contacts as well as the comparison of the size of the real crystals with EIS for different geometry used in our Hall effect measurements. The typical dimensions of our \TTF\ crystals were $4.7\times 0.7\times 0.08\mun{mm}^3$ for the \vb, \va\ and \vc\ direction, respectively. Considering first the  geometry used in  \cite{R08} (\jb, \Ba\ -- cf.\ figure \ref{Fig1}) the EIS coordinate transformation gives the ratio $L'/W' =\ell'_b/\ell'_c = 6.3$,  which is the acceptable value. However the anisotropy is temperature dependent; the ratio  $\sigb/\sigc$ increases from about $10^2$ at room temperature to almost $10^3$ at $60\mun{K}$, and therefore the related $L'/W'$ values  are also changing with temperature. For our sample dimensions this gives at $60\mun{K}$   $\ell'_b/\ell'_c \approx 1.8$, implying that this geometry shows pronounced deterioration with cooling and as a consequence could yield potentially inaccurate results. Indeed, we have obtained very poor results for this geometry. This was also true for the  \jb, \Bc\ geometry (that was used in \cite{R09}) which, for our samples,  already at room temperature had an unsatisfactory value  $L'/W' =\ell'_b/\ell'_a \approx 0.2$, and get even worse with cooling.  These obstacles motivated us to search for yet another geometry that connotes another current direction than \jb\ and that fulfils the $L'/W' \geq 3$ condition for Hall effect measurements. The choice of the sample geometry is however limited with the sample dimensions and disposable area for contacts. The only choice turned out to be  \ja\  because it is impossible to position the Hall contacts along the tiny \vc\ direction. We have also  selected \Bc\ that ensures a remarkable $L/W$ value. Finally, in order to decrease $\ell_b$  and consequently  $\ell'_b$ we have cut the sample along \vb\ direction to approximately $1/4$ of its original dimension. Thus the final dimensions of one of the measured samples  which showed the best Hall effect results, were $1.01 \times 0.88 \times 0.07\mun{mm}^3$ for \vb, \va\ and \vc\ directions, respectively,  with $L'/W' =\ell'_a/\ell'_b = 24$ at room temperature that increases up to $64$ at $60\mun{K}$ (the real crystal and corresponding EIS are also shown on figure \ref{Fig1}). Summing up this part, we point out that our analysis indicate that the  \ja, \Bc\ geometry should be markedly favourable for \TTF\ Hall effect measurements.

\section{Results}

\begin{figure}
\centering
 \includegraphics*[width=\mysize]{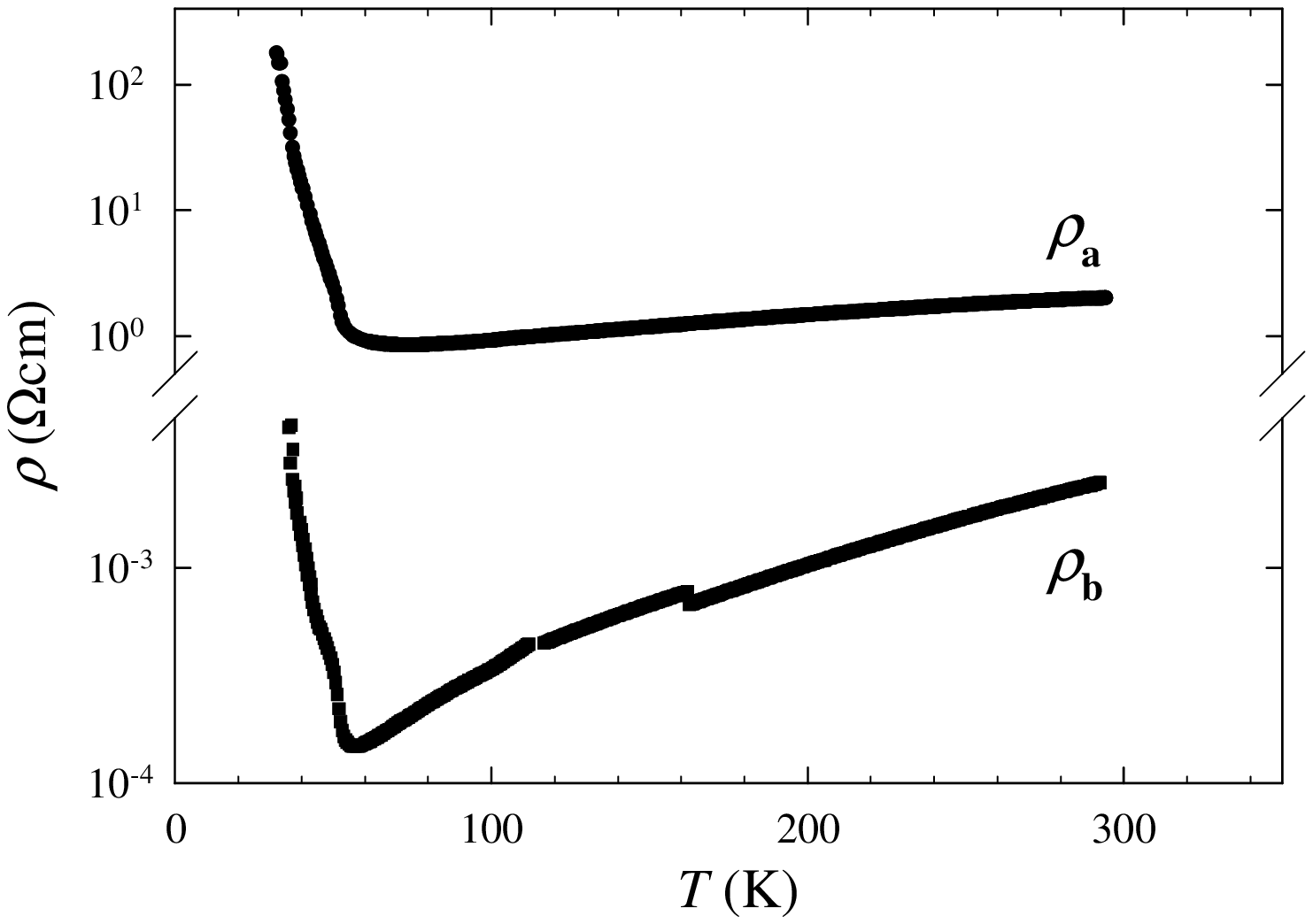}
   \caption{The temperature dependence of the resistivities $\rhob$ and $\rhoa$ (measured along the \vb\ and \va\ crystal directions) for \TTF\  in the temperature range  $30\mun{K}<T<300\mun{K}$. The results are for the samples used for Hall effect measurements.} \label{Fig2}
\end{figure}

Figure \ref{Fig2} shows the temperature dependence of the resistivity along the highest conductivity direction $\rhob(T)$ and along the lowest conductivity direction $\rhoa(T)$, in the temperature range $30\mun{K}<T<300\mun{K}$, measured on samples that were used for Hall effect. The results show  a good agreement with the previously published data \cite{R03} comprising the values of room temperature resistivity, the metallic behaviour for both directions from room temperature down to about $60\mun{K}$ below which the increase of resistivity indicates a cascade of phase transitions (at $54\mun{K}, 49\mun{K}, 38\mun{K}$) that destroy the metallic character.  Further below the phase transitions the resistivity increases exponentially.

\begin{figure}
\centering
 \includegraphics*[width=\mysize]{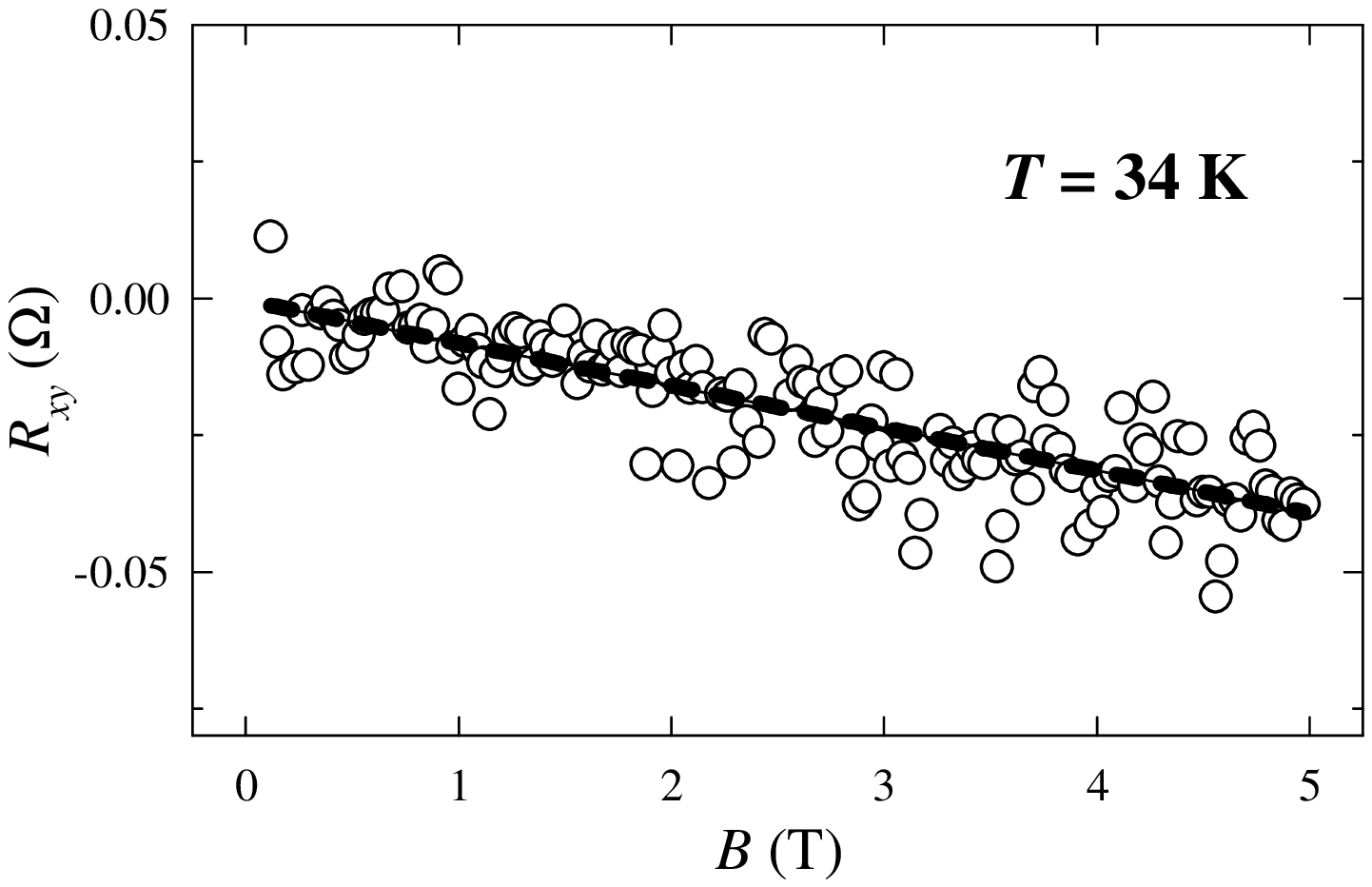}
   \caption{The Hall resistivity vs.\ magnetic field, for \ja, \Bc\ and for $T=34\mun{K}$.} \label{Fig3}
\end{figure}

The magnetic field dependence of the Hall resistance (up to $B = 5\mun{T}$) is shown in figure \ref{Fig3} for $T = 34\mun{K}$ and for \ja, \Bc\ geometry, which we consider as an optimal one. The data show that the Hall resistivity is linear with magnetic field.

\begin{figure}
\centering
 \includegraphics*[width=\mysize]{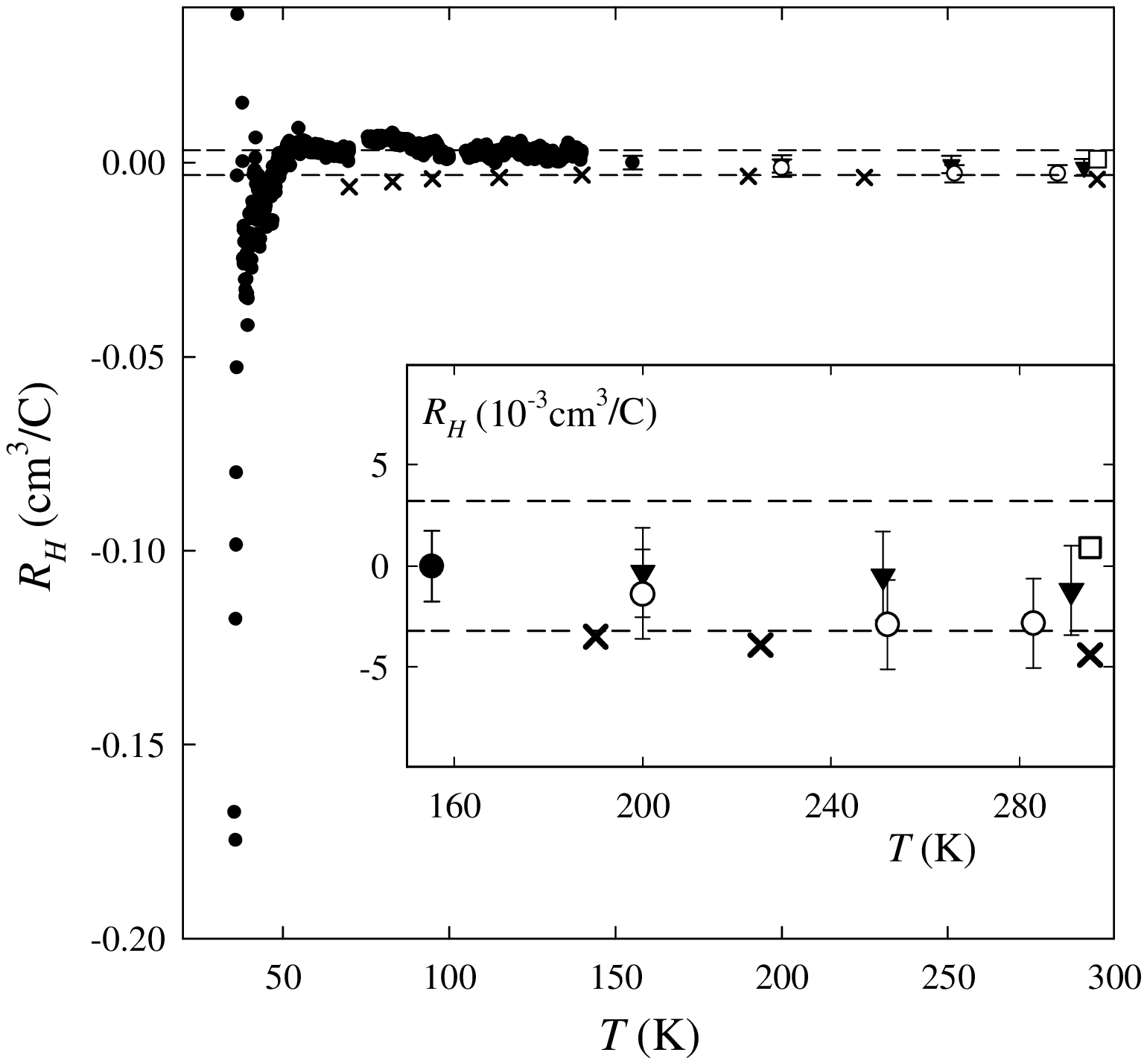}
\caption{The temperature dependence of the Hall coefficient $R_H$ for $30\mun{K}<T<300\mun{K}$. The results are shown for (i) \ja, \Bc, $B=5\mun{T}$, ($\bullet$), the error bars which do not exceed 5\% are not shown; (ii) \jb, \Bc, $B=9\mun{T}$, ($\blacktriangledown$); and (iii) \jb, \Ba, $B=9\mun{T}$, ($\circ$). The results from  \cite{R08} (\jb, \Ba) ($\times$) and \cite{R09} (\jb, \Bc) ($\square$) are also shown. The dashed lines correspond to value calculated for 1D band picture, $R_H = \pm 3.2 \times 10^{-3}\mun{cm}^3$/C -- see section \ref{sec:discussion}. Inset: $R_H(T)$ vs. $T$ for $T>150\mun{K}$ in more details in order to demonstrate $R_H(T) \approx 0$.}\label{Fig5}
\end{figure}

Figure \ref{Fig5} shows the temperature dependence of the Hall coefficient $R_H$ for $30\mun{K}<T<300\mun{K}$. The results are shown for several samples and for different geometries (i) \jb, \Ba; (ii) \jb, \Bc; and (iii) \ja, \Bc. At the increased scale (figure \ref{Fig5} -- Inset) it can be clearly seen that $R_H$ is around zero for $T>150\mun{K}$. Below this temperature, $R_H$ is becoming positive down to phase transitions temperatures region within which it changes its sign  to negative values. It should be noted that for the first two geometries  we have obtained quite scattered data at high temperatures and the signals were not at all detectable at lower temperatures. However, for the \ja, \Bc\ geometry, where the Hall voltage $V_H$ develops along the \vb\ axis,  we have obtained very good results for $T<150\mun{K}$. These measurements were performed on two samples and one of them is presented on figure \ref{Fig5}. Although both samples were good quality ones, we show the data for the sample for which  the Hall contacts were positioned strictly opposite to each other. This is, true enough, a textbook requirement for a good Hall effect measurement, but, due to the extremely small dimensions for this geometry,  not at all an easy task to achieve. The results for another sample confirm the obtained data with positive $R_H$ for $T<150\mun{K}$ and above phase transitions, and negative $R_H$ below.

The data for $T>150\mun{K}$ were determined as the average of the several measurements taken at fixed temperature, during the cooling and heating cycles, using several samples and Hall contacts. The error bars show the uncertainty in the determination of small $R_H$ values. The scattering in the data comes also from the fact that during a field sweeps from $-B_{\mathrm{max}}$ to $+B_{\mathrm{max}}$ at fixed temperatures the small temperature drift occurs (a common and well known source of error in this kind of measurements). Figure \ref{Fig5}  also contains the data from  \cite{R08} and \cite{R09}.

\section{Discussion}
\label{sec:discussion}

As mentioned above, \TTF\ shows metallic behaviour down to about $60\mun{K}$ and undergo the phase transitions at $\THH = 54\mun{K}$, $\TI = 49\mun{K}$ and $\TL = 38\mun{K}$ toward an insulating ground state \cite{R03,R04}. The phase transition at $54\mun{K}$ is manifested as a drop of the conductivity by a factor of about 2 and at $\TL = 38\mun{K}$ a first--order transition toward an insulating ground state occurs. Between $54\mun{K}$  and $38\mun{K}$, charge density waves  (CDWs) successively develop first in the TCNQ and than in the TTF stacks. These transitions have been ascribed to the instability of the one--dimensional electronic gas due to the Peierls mechanism. All CDW phase transitions of \TTF, investigated in detail by X-ray diffuse scattering experiments \cite{R12} and elastic neutron scattering \cite{R13}, have been recently identified using a low temperature scanning tunnelling microscope (STM) \cite{R14} that provides a direct experimental proof for the existence of phase--modulated and amplitude--modulated CDWs.

The present Hall effect investigations of a \TTF\ single crystals were performed in a broad temperature range $30\mun{K}-300\mun{K}$. Although the primary goal was to achieve the best possible experimental conditions in order to provide information regarding the variance of the old Hall effect results (concerning the sign and value of the Hall coefficient at high temperatures above phase transitions),  even more important  was whether the change in the Hall coefficient could be observed in the temperature range of the  phase transitions, which would then give an  experimental proof about successive development of CDWs first in the TCNQ and than in TTF stacks. In what follows we shall therefore discuss separately the obtained data for $R_H(T)$ (regarding the sign and value) for the high ($150\mun{K}<T<300\mun{K}$) and low ($30\mun{K}<T<150\mun{K}$) temperature ranges.

Taking into consideration the quasi--one dimensionality of \TTF, the standard expression for the low field Hall coefficient for an anisotropic tight binding band should be used: $R_H = (1/ne)(k_F b/\tan k_Fb)$  where $b$ is lattice constant along the \vb\ axis. As a partial charge is transferred from TTF to TCNQ stacks and the charge density $\rho$ potentially available for transport is determined by the value of $k_F$ at which the bonding TCNQ band crosses the antibonding TTF band, leading to $2k_F=\rho\pi /b$ in the 1D band picture, we obtain $k_Fb= 0.29\pi$ (taking a band filling of $0.58\mun{el/TCNQ}$) \cite{R12}. The calculated Hall coefficient value is then $R_H=\pm 3.2\times10^{-3}\mun{cm}^3/$C and is represented by dashed line in  figure \ref{Fig5} (`$-$' for electrons only, and `$+$' for holes only). The old data from  \cite{R08} are not far from this calculated value for electrons only. However, our results for all configurations used and for $T>150\mun{K}$ are around $R_H = 0$ (the tiny inclination towards negative values above 250~K is probably irrelevant), indicating that the Hall coefficient results do not depend on the geometry conditions during measurements. If our data are compared with those from \cite{R08} (where the negative Hall coefficient was obtained) and \cite{R09} (where the positive Hall coefficient was obtained for another geometry), apparently, the difference could not be solely attributed to the geometry configuration used for measurements. It would be correct to note, that the possible cause could be also the different sample quality, which is now difficult to confirm, but cannot be excluded. Namely, the thermoelectric power (TEP) results (from the same period as \cite{R08,R09}) have shown that at high temperatures the TEP is negative and approximately linear in temperature indicating the metallic state, while at $T < 54\mun{K}$ positive and negative signs were found in samples from different sources \cite{R17a,R17b}.  On the other hand, our results are quite consistent with the measurements of $^{13}$C NMR in \TTF\ (obtained more than 20 years ago as well \cite{R15}) that were somehow in contradiction with expected `narrower' TTF band due to the previous suggestions that the dominant carriers are electrons located on TCNQ stacks \cite{R08}. In conclusion of this part, we state that our Hall coefficient results, above 150~K,
show $R_H \approx 0$ for all geometries studied, and do not confirm dominance of neither
electrons nor holes in transport properties \cite{R16}. In other words, they show the possibility
that electron--like (TCNQ stacks) and hole-like (TTF stacks) contributions effectively
cancel each other giving $R_H \approx 0$, which indicate that both chains equally contribute to
the conductivity.

\begin{figure}
\centering
 \includegraphics*[width=\mysize]{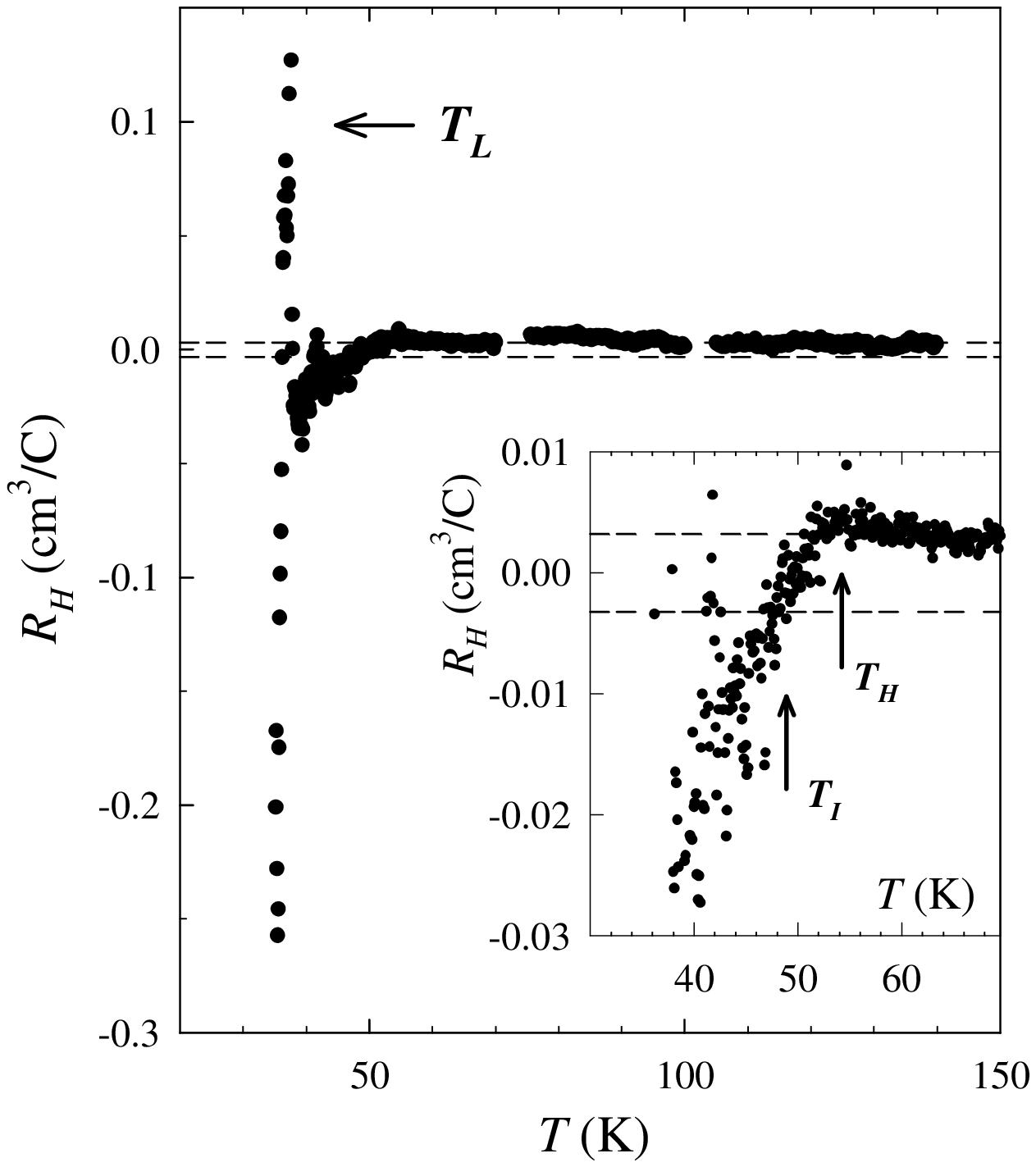}
\caption{The temperature dependence of the Hall coefficient $R_H(T)$ between $30\mun{K}$ and $150\mun{K}$ for \ja, \Bc, for our best sample (the error bars which amount to around 5\% are not shown). The strong upturn in $R_H(T)$ indicates the phase transition $\TL = 38\mun{K}$. Inset shows the temperature interval around phase transitions $\THH = 54\mun{K}$ and $\TI = 49\mun{K}$. The dashed lines correspond to values calculated for 1D band picture, $R_H = \pm 3.2 \times 10^{-3}\mun{cm}^3$/C -- see text}\label{Fig7}
\end{figure}

The temperature dependence of the Hall coefficient in the $30\mun{K}<T<150\mun{K}$ temperature range for our best sample and for the \ja, \Bc\ geometry is presented in figure \ref{Fig7}. As discussed previously, the other two geometries (\jb, \Ba\ and \jb, \Bc) did not give conceivable data due to the geometry limitation for anisotropic samples. For $100\mun{K}<T<150\mun{K}$, $R_H$ is small and positive indicating that TTF stack (holes) contributes more to the conductivity than TCNQ stack (electrons). Here, we would like to remind again  on the  $^{13}$C NMR measurements \cite{R15}, which have shown that the density of states of TCNQ stacks starts decreasing appreciably below about $150\mun{K}$: this  may explain the fact that $R_H$ was not detectable above $150\mun{K}$ indicating the equal contribution of TTF and TCNQ stacks in the transport, while below $150\mun{K}$ a small positive $R_H$ may be correlated to the decrease of the density of states on TCNQ stacks. We do not give special significance to the obtained $R_H$ values regarding the determination of accurate number of electrons and/or holes that participate in transport properties, as there is substantial experimental evidence that Coulomb interaction plays an essential role in the electronic structure of \TTF\ (\cite{R04,R07a,R07b,R07c}) and a purely band theoretical description may be inadequate. The smooth increase of positive $R_H$ below $100\mun{K}$ down  to  $\THH \approx 54\mun{K}$ -- which  is clearly perceived in the inset of figure \ref{Fig7} -- can also be ascribed to a pre-transitional behaviour. However, as the temperature is lowered in the phase transition range, our Hall data  show features which have not been observed up to now: all three phase transitions of TTF-TCNQ are identified from the Hall effect measurements as indicated on figure \ref{Fig7}.  Around $\THH = 54\mun{K}$ there is a maximum in positive $R_H(T)$ that decreases with further cooling and changes its sign around $\TI = 49\mun{K}$. The strong temperature dependence of $R_H(T)$ (a pronounced upturn) close to the phase transition at $\TL = 38\mun{K}$ marks the first-order phase transition (where, as known, the period in the $a$ direction jumps discontinuously to $4a$ \cite{R04}) toward insulating ground state. Here it is worth pointing out that such $R_H(T)$ result was not foreseen. Namely, the phase transition at 54~K is driven by the CDW Peierls instability in the TCNQ chains, that is manifested in a drop of the conductivity by a factor of about 2 \cite{R04}. We have expected that in the temperature region $\TL<T<\THH$ the dominance of the TTF stacks (i.e. the hole-like carriers contribution to the transport) in Hall coefficient measurements may be found. The gradual decrease of positive $R_H$ below $\THH$ (where the reduction of the number of electron-like carriers is expected) and the observed change of sign around $\TI$ suggest that both kind of carriers contribute to the conductivity in this transitional region. For the Hall coefficient in a two-band model  the resultant value and the sign are determined by the carriers concentrations  as well as their mobilities  that both have strong and probably different temperature dependencies in this temperature region. $R_H(T)$ then comes from a balance  of the hole and electron terms \cite{Mott}. These facts are even more pronounced with the well-marked upturn close to the phase transition at $\TL = 38\mun{K}$. This kind of behaviour is a known feature in semiconductors \cite{R16}. As the Hall fields created by electrons and holes are opposing each other, the galvanomagnetic effects can have unusually strong temperature variations in regions where the resultant Hall field is nearly zero and where the relative electron-hole population is temperature dependent. Below phase transitions, for $T<38\mun{K}$, $R_H(T)$ is negative with the value that rapidly increases with further temperature decrease following the activated behaviour that corresponds to that of resistivity.

\section{Conclusion}
In summary, we have reported measurements of the Hall coefficient $R_H(T)$ at ambient pressure of the quasi--one dimensional organic conductor \TTF\ for $30\mun{K}<T<300\mun{K}$ and in magnetic fields up to $9\mun{T}$.  We have applied the equivalent isotropic sample approach, suggested the way  of choosing the best geometry for Hall effect measurements of highly anisotropic samples, and proposed that for the \TTF\ system  the  \ja, \Bc\ configuration is the best choice.

Above $150\mun{K}$, we could not confirm the dominance of neither electrons nor holes in the transport properties and we suggest the possibility that both chains, electron--like (TCNQ stacks) and hole--like (TTF stacks), equally contribute to the conductivity giving $R_H(T) \approx 0$. Also, based on our data for other measurement geometries, we indicate that the differences in geometry are not important regarding the sign and/or value of the Hall coefficient in this temperature range. For $T<150\mun{K}$ our results show small and positive $R_H(T)$ that may indicate the dominance of holes in this region, and an increase below $100\mun{K}$ down to phase transition at $\THH=54\mun{K}$ that could be ascribed to a pre-transitional behaviour. For
$30\mun{K}<T<54\mun{K}$ temperature region we have shown that
all three phase transitions of TTF-TCNQ can be identified from the Hall effect measurements:  (i) around $\THH = 54\mun{K}$ there is a maximum in positive $R_H(T)$  (ii)  around $\TI = 49 K$, $R_H(T)$ changes sign becoming negative for lower temperatures (iii) around $\TL = 38\mun{K}$, a pronounced upturn in $R_H(T)$ marks the first-order
phase transition towards the insulating ground state for
$T<\TL$ that is characterized by activated $R_H(T)$ with the
activation energy corresponding to that of resistivity.

\ack
The work was supported by the Croatian Ministry of Science, Education and Sports grants 119--1191458--1023 and 035--0000000--2836.

\section*{References}

\bibliography{etafra}

\end{document}